\documentclass[12pt]{article}
\usepackage[affil-it]{authblk}
\usepackage{amsmath}
\usepackage{amssymb}
\usepackage[a4paper]{geometry}
\usepackage{amsthm}
\usepackage{subcaption}
\usepackage{graphicx}
\usepackage{tikz}
\usepackage{tikz-cd}
\usepackage{mathrsfs}
\usepackage{bm}
\usepackage{bbold}
\usepackage{slashed}
\usepackage[hidelinks]{hyperref}
\usepackage{cancel}
\hypersetup{
    colorlinks = true,
    linkcolor = {blue},
    citecolor = {blue},
    anchorcolor = {green},
    citecolor = {red},
    filecolor = {cyan},
    menucolor = {red},
    runcolor = {cyan},
    urlcolor = {magenta}
}
\newcommand{\twidth}{6in}
\textwidth=\twidth
\usepackage{amsmath,amssymb, braket, lscape, graphicx, comment}

\newcommand{\II}{{\mathcal{I}}}
\newcommand{\LL}{{\mathcal{L}}}
\newcommand{\CC}{{\mathcal{C}}}

\newcommand{\XX}{{\mathcal{X}}}
\newcommand{\OO}{{\mathcal{O}}}
\newcommand{\UU}{{\mathcal{U}}}
\newcommand{\VV}{{\mathcal{V}}}
\newcommand{\MM}{{\mathcal{M}}}
\newcommand{\PP}{{\mathcal{P}}}

\newcommand{\su}{{\mathfrak{su}}}

\renewcommand{\O}{{\mathrm{O}}}
\newcommand{\SU}{{\mathrm{SU}}}
\newcommand{\SO}{{\mathrm{SO}}}

\newcommand{\A}{{\mathbb{A}}}
\newcommand{\R}{{\mathbb{R}}}

\newcommand{\Z}{{\mathbb{Z}}}

\newcommand{\C}{{\mathbb{C}}}
\renewcommand{\H}{{\mathbb{H}}}

\newcommand{\1}{{\mathbb{1}}}

\newcommand{\bu}{{\bm{1}}}

\newcommand{\bi}{{\bm{i}}}
\newcommand{\bj}{{\bm{j}}}
\newcommand{\bk}{{\bm{k}}}
\newcommand{\balp}{\boldsymbol{\alpha}}
\newcommand{\beq}{\begin{equation}}
\newcommand{\eeq}{\end{equation}}
\newcommand{\bea}{\begin{eqnarray}}
\newcommand{\eea}{\end{eqnarray}}
\newcommand{\bal}{\begin{align}}
\newcommand{\eal}{\end{align}}
\newcommand{\bml}{\begin{multline}}
\newcommand{\eml}{\end{multline}}

\newcommand{\bdy}{\partial}

\newcommand{\wt}{\widetilde}

\newcommand{\lto}{\longrightarrow}

\def \d{\mathrm{d}}

\newcommand{\tr}{{\operatorname{tr}}}

\newcommand{\id}{{\rm Id}}

\newcommand{\diag}{{\rm diag}}

\makeatletter
\newcommand\xleftrightarrow[2][]{%
  \ext@arrow 9999{\longleftrightarrowfill@}{#1}{#2}}
\newcommand\longleftrightarrowfill@{%
  \arrowfill@\leftarrow\relbar\rightarrow}
\makeatother

\newcommand{\bigslant}[2]{{\raisebox{.2em}{$#1$}\left/\raisebox{-.2em}{$#2$}\right.}}

\title{Quantization of skyrmions using instantons}
\author{{\Large Josh Cork}%
  \thanks{Email address: \texttt{josh.cork@leicester.ac.uk}}}
\affil{{School of Computing and Mathematical Sciences, University of Leicester,}\\
University Road, Leicester, LE1 7RH, United Kingdom}

\author{{\Large Chris Halcrow}%
  \thanks{Email address: \texttt{chalcrow@kth.se}}}
\affil{Department of Physics, KTH-Royal Institute of Technology,\\
 Stockholm, SE-10691 Sweden}
\date{\large \today}
\numberwithin{equation}{section}

\begin{document}
\renewcommand*{\thefootnote}{\fnsymbol{footnote}}
\begin{titlepage}
\begin{center}
{\LARGE Quantization of skyrmions using instantons \par}


\vspace{10mm}
{\Large Josh Cork\footnote{Email address: \texttt{josh.cork@leicester.ac.uk}}$^{\rm 1}$\ and Chris Halcrow$^{\rm 2}$}\\[10mm]

\noindent {\em ${}^{\rm 1}$ School of Computing and Mathematical Sciences\\
University of Leicester, University Road, Leicester, United Kingdom
}\\
\smallskip
\noindent {\em ${}^{\rm 2}$ Department of Physics, KTH-Royal Institute of Technology,\\
 Stockholm, SE-10691 Sweden}\\[10mm]
{\Large \today}
\vspace{15mm}
\begin{abstract}
    We provide a step-by-step method to construct skyrmions from instanton ADHM data, including when the exact ADHM data is unknown. The configurations look like clusters of smaller skyrmions, and can be used to build manifolds of skyrmions with or without symmetries. Nuclei are described by quantum states on these manifolds. We describe the construction and quantization procedure generally, then apply the methods in detail to the 8-skyrmion which describes the Beryllium-8 nucleus.
\end{abstract}
\end{center}
\end{titlepage}
\renewcommand*{\thefootnote}{\arabic{footnote}}
\setcounter{footnote}{0}
\hypersetup{
    linkcolor = {blue}
}
\section{Introduction}

Skyrmions are topological solitons used to model nuclei \cite{skyrme1962nucl}. Each classical configuration has a topologically protected integer, $N$, which is identified with the baryon number of the corresponding nucleus. To compare properties such as energies, allowed (iso)spins, and charge radii one must quantize the classical skyrmion. This may be done semiclassically by selecting a low energy manifold of configurations, which we’ll call the configuration space, and solving a Schrödinger equation on it. In the simplest case one chooses the minimal energy skyrmion and its zero modes: translations, rotations, and isorotations \cite{AdkinsNappiWitten1983static, Irwin:1998bs}. But it is well known that skyrmions and nuclei deform. Hence to improve the approximation one should include deformations in their configuration space. This is difficult due to the nonlinearity of the Skyrme model. Previous studies which attempt to include deformations are limited in different ways. They: generated one-dimensional submanifolds using gradient flow \cite{Leese:1994hb, Halcrow:2015rvz}, made phenomenological guesses about the space \cite{Halcrow:2016spb}, or used a harmonic approximation \cite{Walet:1996he, gudnasonhalcrow2024quantum}.

In this paper we’ll develop a method that uses the instanton approximation, based on ADHM data, to construct configuration spaces in the Skyrme model; as an example we apply this method to the quantization of the 8-skyrmion. Atiyah and Manton first suggested that instantons, soliton solutions of Yang--Mills theory, could be used to approximate skyrmions \cite{AtiyahManton1989}.  The idea is powerful as instantons are incredibly well understood. In each topological sector, there are 8$N$ instantons and all can be described by $(N+1)\times N$ quaternionic matrices, called ADHM data \cite{ADHM1978construction}.

There has been significant recent progress in understanding the link between skyrmions and instantons. The Atiyah--Manton approximation is understood by viewing the Skyrme lagrangian as the first term in an expansion relating instantons to skyrmions coupled to vector mesons \cite{sutcliffe2010skyrmions}, and this has been used to interpret modes in 2-skyrmion scattering \cite{Halcrow:2021wwc}. A new numerical method to generate skyrmions from instantons was developed in \cite{CorkHarlandWinyard2021gaugedskyrmelowbinding,harland2023approximating}; this has been used to construct larger spaces of skyrmions from instantons than ever before \cite{corkhalcrow2022adhm}, and to generate explicit rational approximations of skyrmions \cite{harlandsutcliffe2023rational}. Very recently, a new formula for calculating Finkelstein--Rubinstein constraints directly from ADHM data has been found \cite{corkharland2024FR} improving on old results which were only applicable to rigid body quantization \cite{krusch2003homotopy}.

Despite this progress, the instanton approximation is still not widely used. Hence one aim of this paper is to write a simple ``recipe” to generate ADHM data, and hence skyrmion spaces, quickly and easily. After briefly reviewing the Skyrme model, Atiyah--Manton construction, and ADHM data in Section \ref{sec.review}, we outline such a step-by-step guide in Section \ref{sec:gen-method}, along with a review of the quantization procedure. These steps are applied in detail in Section \ref{sec:8-sky}, and some more detailed (but more generic) examples are showcased in Section \ref{sec:paths-no-symm}.

\section{Skyrmions, instantons, and ADHM data}\label{sec.review}

The lagrangian density of the Skyrme model, with zero pion mass, is given by
\begin{equation} \label{eq:SkyrmeLagrangian}
\mathcal{L} = -\frac{f_\pi^2}{16\hbar}\text{tr} \left( L_\mu L^\mu \right) + \frac{\hbar}{32e^2}\text{tr}\left( [L_\mu, L_\nu][L^\mu, L^\nu ] \right) ,
\end{equation}
and is written in terms of the left-invariant current $L=U^{-1}\d U$, for $U:\R^3\lto \SU(2)$. For simplicity, we choose energy and length units $f_\pi/4e$ and $2\hbar/f_\pi e$ respectively; these are called Skyrme units. Skyrmions are energy-minimising maps $U:\R^3\lto\SU(2)$ of the dimensionless static Skyrme energy
\begin{align}\label{Skyrme-energy}
    E(U)=-\frac{1}{2}\int\tr\left(L_iL_i+\tfrac{1}{8}[L_i,L_j][L_i,L_j]\right)\,\d^3x,
\end{align}
satisfying the space-compactifying boundary condition $U\to\id$ as $r\to\infty$; this boundary condition allows for a well-defined topological degree $N\in\Z=\pi_3(\SU(2))$, physically identified as the baryon number, and computed via the integral
\begin{align}\label{baryon-number}
    N=\frac{1}{24\pi^2}\int\tr(L\wedge L\wedge L).
\end{align}
The energy is bounded below proportionally by the charge \cite{faddeev1976some}, and the choice of units has been made so that the energy bound is $E\geq 12\pi^2|N|$.

The Euler--Lagrange equations for \eqref{Skyrme-energy} are highly nonlinear and no solutions are known analytically. A good approximation of solutions is provided by instantons on $\R^4$ \cite{AtiyahManton1989}. Instantons are gauge fields $A$ on $\R^4$ with anti-self-dual curvature $\star F=-F$ which extend smoothly to the one-point compactification $S^4\cong\R^4\cup\{\infty\}$. This boundary condition identifies each instanton with a topological charge $N\in\Z$ called the instanton number, computed as the second Chern number
\begin{align}
    N=c_2(S^4)=\frac{1}{8\pi^2}\int\tr(F\wedge F).
\end{align}
For each $N$ there is an $8|N|$-dimensional moduli space $\II_N$ of instantons modulo gauge transformations which tend to identity at infinity \cite{atiyahhitchinsinger1978self}. The moduli spaces $\II_N$ are parameterised by a moduli space of matrices called ADHM data \cite{ADHM1978construction}. The ADHM data may be described as follows. Let $\XX_N$ denote the set of all pairs $(L,M)$ where $L$ is a length-$N$ row vector of quaternions, and $M$ is an $N\times N$ symmetric matrix of quaternions. For any element $(L,M)\in\XX_N$, and $x\in\H$, one may write down the associated matrix
\begin{align}
    \Delta_x=\begin{pmatrix}
        L\\
        M-x\id_N
    \end{pmatrix}.
\end{align}
To describe an instanton at $x=x_1\bi+x_2\bj+x_3\bk+x_4\bu\in\H\cong\R^4$, these data must satisfy the reality condition: that the $N\times N$ matrix $\Delta_x^\dagger\Delta_x$ is real and invertible for all $x\in\H$. The moduli space $\A_N$ of ADHM data is the set of all $(L,M)\in\XX_N$ satisfying the reality condition, modulo the action of $\O(N)$ given by
\begin{align}
    O\cdot(L,M)=(LO^{-1},OMO^{-1}),\quad O\in\O(N).
\end{align}
The instanton associated to any ADHM data is given by an induced connection on $\ker\Delta^\dagger$; the subbundle of the trivial quaternionic bundle $\R^4\times\H^{N+1}$ with fibers $\ker\Delta_x^\dagger$. Explicitly, one solves for each $x\in\H$
\begin{align}
    \Delta_x^\dagger\Psi_x=0,\quad \Psi_x^\dagger\Psi_x=\1,\quad\Psi_\infty=\begin{pmatrix}
        \bu&0&\cdots&0\end{pmatrix}^t,
\end{align}
and sets $A|_x=\Psi_x^\dagger\d\Psi_x$.

The Atiyah--Manton approximation of skyrmions \cite{AtiyahManton1989} generates an approximate skyrmion $U:\R^3\lto\SU(2)$ as the holonomy of an instanton along all lines of fixed imaginary part $\Im(x)\in\Im(\H)\cong\R^3$. In this scheme, the instanton number $N$ is the baryon number of the associated Skyrme field, hence why we use the same symbol for both quantities. In general, it is not possible to write down instanton holonomies analytically, so one must approximate them numerically. Since instantons are understood via ADHM data as induced connections, we may approximate the holonomy using the methods outlined in \cite{harland2023approximating}. In this paper, we approximate the holonomy using the improved order 3 method, the details of which may be found in \cite{harland2023approximating}.

\section{General methodology}\label{sec:gen-method}

Previous work has focused on finding the ADHM data that has the same, often large, symmetry group as a known skyrmion. We will generalize this problem and try to construct ADHM data which looks like $k$-clusters of smaller skyrmions, which can have any symmetry. These will often be families of data, used to generate a configuration space of skyrmions. In this section we shall discuss how to generate ADHM data of this type. We will also discuss a semiclassical quantization on a configuration space.

\subsection{Constructing clustered configurations}\label{sec:cluster-method}

To describe any configuration in terms of smaller clusters, we follow a general road map outlined below. The procedure, formulated through the framework introduced in \cite{corkhalcrow2022adhm}, will generate ADHM data, hence a genuine instanton, which models the skyrmion configuration of interest.
\paragraph{Step 1:} Understand the component clusters and their ADHM data. These will be $k$ true ADHM data $(L_i,M_i)\in\A_{N_i}$, $i=1,\dots,k$, which describe the constituent components. A list of ADHM data describing common components, such as tori and cubes, may be found in \cite{corkhalcrow2022adhm}.
\paragraph{Step 2:} Write down specific `test data' $(L_T,M_T)\in\XX_N$ (with $N=\sum_iN_i$) which describes the problem. The test data is explicitly given by embedding each constituent of $\A_{N_i}$ into $\XX_N$ as a block diagonal:
\begin{align}\label{generic-test-data}
    \begin{aligned}
        L_T&=\begin{pmatrix}
            p_1L_1q_1^{-1}&\cdots&p_kL_kq_k^{-1}
        \end{pmatrix},\\
        M_T&=\diag\{q_1M_1q_1^{-1}+a_1\id_{N_1},\dots,q_kM_kq_k^{-1}+a_k\id_{N_k}\}.
    \end{aligned}
\end{align}
The test data \eqref{generic-test-data} will generally not be ADHM data as it will not solve the reality condition; it is instead a building block from which to determine true ADHM which describes the physical system of interest. In \eqref{generic-test-data} the $(p_i,q_i)\in\SU(2)^2$ are unit quaternions that describe the orientations of the components in isospace and space respectively, and the $a_i\in\H$ describe the positions. Typically one sets $\Re(a_i)=0$ so then the $a_i$ are $3$-vectors represented by imaginary quaternions; separations in the real (holographic) direction may be relevant for parameterising vector meson modes \cite{Halcrow:2021wwc}, but we do not consider this possibility here.

If the system requires any inverted constituents, i.e. those whose Skyrme fields differ by the parity transformation $U(x)\mapsto U(-x)^{-1}$, one simply replaces the relevant constituent data $(L_i,M_i)$ by $(-L_i,-M_i)$.
\paragraph{Step 3:} Determine the symmetry of the entire system. The group $\SU(2)\times\SU(2)\times\Z_2$ of isorotations, rotations, and parity transformations, acts on $(L,M)\in\XX_N$ via
\begin{align}
\begin{aligned}
    (p,q,1)\cdot(L,M)&=(pLq^{-1},qMq^{-1}),&(p,q,-1)\cdot(L,M)&=(-pLq^{-1},-qMq^{-1}).
\end{aligned}
\end{align}
The test data $(L_T,M_T)$ may be invariant, up to gauge transformation, under a subgroup of these: this is the stabiliser subgroup $S_T$ of \eqref{generic-test-data} in $\SU(2)\times\SU(2)\times\Z_2$. A detailed discussion of identifying such symmetries may be found in \cite{corkhalcrow2022adhm}. In brief, the symmetry is made explicit by finding, for each generator $g\in S_T$ of the symmetry group of \eqref{generic-test-data}, a compensating gauge transformation $O_g\in\O(N)$ such that
\begin{align}\label{symmetry-test}
    (L_T,M_T)=g\cdot(L_TO^{-1}_g,O_gM_TO_g^{-1}).
\end{align}
\paragraph{Step 4:} Construct the most general $(L,M)\in\bigslant{\XX_N}{\O(N)}$ consistent with the system and symmetry. Starting with the rigid body symmetries $S_T$, one may write down the most general $(L,M)\in\XX_N$ with these symmetries, i.e. such that
\begin{align}\label{general-symm}
    (L,M)=g\cdot(LO^{-1}_g,O_gMO_g^{-1})\quad\text{for all }g\in S_T,
\end{align}
where $O_g$ are the compensating gauge transformations identified earlier in \eqref{symmetry-test}. Our approach here contrasts with previous work (for example in \cite{AllenSutcliffe2013,FurutaHashimoto1990}) which found symmetric ADHM data by fixing a symmetry group $G$ and performed an exhaustive search for all ADHM data with that symmetry. In general there will be several choices of compensating gauge transformations which allow for solutions of \eqref{general-symm} given by different $N$-dimensional representations of $G$, and the fixed point set under the action of $G$ may well be disconnected in $\A_N$, with some different representations describing different components. In our case the compensating gauge transformations are fixed in Step 3, and this guarantees we end up with data in the connected component of the fixed point set of $S_T$ in $\XX_N$ which describes the physical system of interest.

Even after imposing symmetry, the most general $(L,M)\in\XX_N$ found in this way will likely depend on several free parameters. Some of these will be fixed by the reality condition in the next step, whereas others may be redundancies due to gauge freedom and may be removed. Note that \eqref{generic-test-data} acts as a partial gauge fixing imposed by the choice of gauge for the individual components $(L_i,M_i)$, and this fixes the compensating gauge transformations arising in the symmetry. To respect this, the residual gauge freedom may be determined explicitly as the subgroup $\OO\subset\O(N)$ which commutes with the compensating gauge transformations, i.e.
\begin{align}
    \OO=\{\Omega\in\O(N)\::\:\Omega O_g=O_g\Omega\,\text{ for all }g\in S_T\}.
\end{align}
Finally, the remaining free parameters may be thought of as depending on a set of physical parameters $R_i$ which are prescribed by the parameters of the test data. These may be constrained by internal symmetries of the test data, manifested by one-parameter families $R(t)$, $t\in[0,1]$, such that there exists $\Omega\in\O(N)$ with
\begin{align}\label{internal-symmetry}
    (L_T(R(0)),M_T(R(0)))=(pL_T(R(1))q^{-1}\Omega^{-1},\Omega qM_T(R(1))q^{-1}\Omega^{-1}).
\end{align}
in the chosen gauge.
\paragraph{Step 5:} Find the ADHM data which is closest to $\MM_T(R_i)=(L_T,M_T)(R_i)$ consistent with the symmetries. To do so, start with the general data $(L,M)\in\bigslant{\XX_N}{\O(N)}$ consistent with the symmetries found in Step 4. We'll parameterise this set by $r_j$, and denote the general matrices as $\mathcal{M}(\boldsymbol{r})$. We then try to minimise $||\mathcal{M}(r_j) - \MM_T(R_i)||$, while enforcing the reality condition, which will project $\mathcal{M}({\bm r})$ from $\bigslant{\XX_N}{\O(N)}$ to $\A_N$; the space of ADHM data. Explicitly, we need to solve
\begin{equation}\label{projection}
    \min_{\boldsymbol{r}} || \mathcal{M}\left( \boldsymbol{r}\right) - \mathcal{M}_T\left(R_i \right) || \quad \text{subject to} \quad \Im\left( \mathcal{M}(\boldsymbol{r})\mathcal{M}^\dagger(\boldsymbol{r}) \right) = 0 .
\end{equation}
Throughout, we'll choose the gauge-invariant distance function induced by the standard inner product on matrices
\begin{equation}
\langle A, B \rangle = \tr\left( A^\dagger B \right).
\end{equation}
In general, for complicated clusters, resolving the reality condition (and hence even more so the problem \eqref{projection}) is not possible to do analytically, so we have developed a numerical procedure to overcome this. 

\subsection{Quantization procedure} \label{sec:quantizationprocedure}
In Section \ref{sec:cluster-method} above, we detailed how to generate ADHM data describing any desired physical configuration of Skyrme fields. These data are parameterised by physical parameters $R_i$ which may be thought of as coordinates for a vibrational manifold $\VV$. Upon applying the Atiyah--Manton construction, this family of ADHM data generates Skyrme fields which form a finite-dimensional submanifold $\VV$ of the true configuration space $\CC_N={\rm Maps}_N(S^3,S^3)$. The aim is to quantize the Skyrme model on this manifold coupled with the manifold of zero modes: the group $\SU(2)^2$ of isorotations and rotations. Na\"ively this restricted configuration space is a product manifold $\SU(2)\times\SU(2)\times\VV$, however this may not be true globally. The correct perspective, as pointed out by Rawlinson \cite{Rawlinson:2019xsn}, is to view the restricted configuration space of interest as a principal $\SU(2)^2$ bundle $\PP\to\VV$ over $\VV$. The configuration space on which to quantize the theory is then $\PP$, which is a $(6+\dim(\VV))$-dimensional manifold which locally looks like $\SU(2)^2\times\VV$; formally, there is a surjective map $\pi:\PP\to\VV$, such that for any point $R\in\VV$, there is a neighbourhood $\UU\subset\VV$ of $R$, so that $\pi^{-1}(\UU)\cong\SU(2)^2\times \UU$.
We now briefly review what the quantization procedure on $\PP$ entails. We perform a canonical quantization on this manifold by resolving the Schr\"odinger equation
\begin{equation}\label{eq:basic-Schro}
	\left(-\frac{\hbar^2}{2}\Delta_g + V\right)\Psi = E\Psi,
\end{equation}
where $\Delta_g$ and $V$ are the Laplace--Beltrami operator and potential on $\PP$ inherited from the Skyrme lagrangian. Local formulae for the metric and potential are determined by constructing a family $U_X$ of Skyrme fields dependent on $X\in\SU(2)^2\times\VV$ as
\begin{align}
U_X(\vec{x};X)=pU_R(q\cdot\vec{x})p^{-1},\label{adiabatic-U}
\end{align}
where $U_R$ denotes the family of Skyrme fields parameterised purely by the vibrational coordinates $R\in\VV$, and $p,q$ are general points in $\SU(2)$. Letting these coordinates depend on time, we introduce angular velocity vectors associated to the isorotations and rotations respectively:
\begin{align}\label{angular-vel}
    a_i=-i\tr(\tau_ip^{-1}\dot{p})\quad b_i=i\tr(\tau_i\dot{q}q^{-1});
\end{align}
these give rise to associated right (and left resp.) invariant one forms $\alpha_i$ and $\beta_i$ on $\SU(2)$. Here, as usual, $\tau_i$ denotes the standard (hermitian) Pauli matrices. Inserting \eqref{adiabatic-U} into the dimensionless lagrangian density
\begin{align}
    \LL&=-\frac{1}{2}\int\tr(L_\mu L^\mu-\tfrac{1}{8}[L_\mu,L_\nu][L^\mu,L^\nu])\,\d^3x=T_g-V,
\end{align}
the local metric on $\PP$ is then extracted from the kinetic energy
\begin{align}\label{metric-KE}
\begin{aligned}
    T_g &=\frac{1}{2}g_{ij}\dot{X}_i\dot{X}_j,&
    g_{ij} &= -\int \tr\left( G_i G_j  + \tfrac{1}{4} [L_k, G_i][L_k,G_j] \right) \d^3x,
\end{aligned}
\end{align}
where 
\begin{align}
    \dot{X}_i=\begin{cases}
        a_i,&i=1,2,3,\\
        b_{i-3},&i=4,5,6,\\
        \dot{R}_{i-6},&i\geq7,
    \end{cases}
\end{align}
and $G_i:\R^3\lto\su(2)$ is a current determined by derivatives of \eqref{adiabatic-U} with respect to the coordinates $X_i$. Because of the (iso)rotational symmetry of the lagrangian, the metric terms corresponding to the $\SU(2)^2$ action are generated by special $G_i$ that only depend on the skyrmion at fixed $R\in\VV$:
\begin{align}\label{G-SU2}
    G_i = \begin{cases}  \tfrac{i}{2}U_R^{-1}[ \tau_i, U_R ] \quad &i=1,2,3,\\
    \varepsilon_{(i-3)lm}x_l L_m \quad &i=4,5,6,\end{cases}
\end{align}
where $L_m=U_R^{-1}\bdy_mU_R$. Terms corresponding to $\VV$ explicitly involve derivatives with respect to the parameters $R_i$:
\begin{equation}\label{G-vib}
    G_i = U_R^{-1}(\partial_{R_{i-6}} U_R)\quad i\geq 7 \, .
\end{equation}
Due to the $\SU(2)^2$ action, we can decompose the wave function via an expansion into spin and isospin states as
\begin{align}\label{wave-function-expansion}
    \Psi=\sum_{K_3=-I}^I\sum_{L_3=-J}^J\psi_{K_3,L_3}(R_i)|I,K_3\rangle\otimes|J,L_3\rangle.
\end{align}
Here $I$ and $J$ are isospin and spin, and $K_3$ and $L_3$ are the body-projected isospin and spin respectively. There are also the space-projected isospin and spin $I_3$ and $J_3$ respectively, but these have no effect on the energy spectrum, so we are free to set $I_3=J_3=0$ throughout. In this framework, the wave function is then a function of local coordinates $R_i\in\UU\subset\VV$ on the vibrational manifold taking values in $\C^{(2I+1)(2J+1)}$, with components $\psi_{L_3,K_3}(R_i)$. However, the formula \eqref{wave-function-expansion} is only valid locally, and only extends globally when $\PP$ is trivial. In general the expansion \eqref{wave-function-expansion} gives the local formula for a section $\Psi$ of the associated vector bundle $\PP\times_{\SU(2)^2}
    \C^{(2I+1)(2J+1)}$, in which suitable conditions are imposed on overlapping patches using the transition functions on $\PP$ \cite{Rawlinson:2019xsn}. In each trivialisation the wavefunction \eqref{wave-function-expansion} can be substituted into the Schr\"odinger equation \eqref{eq:basic-Schro}, which becomes a PDE on a manifold with dimension equal to the number of vibrational coordinates $R_i$.

\subsection{Finkelstein--Rubinstein constraints}

Although we are working with a truncated configuration space modeled by instanton moduli, ultimately this acts as an approximation to the true configuration space of Skyrme fields $\CC_N={\rm Maps}_N(S^3,S^3)$. The scheme outlined in the previous section is an approximation to obtaining a Schr\"odinger equation on $\CC_N$. In the untruncated picture, the wave function is really a map\footnote{Again, this is a local picture; in reality $\Psi$ is a section of a complex line bundle over $\CC_N$.} $\Psi:\wt{\CC_N}\lto\C$ defined on the universal cover $\wt{\CC_N}$ of $\CC_N$. Since $\pi_1(\CC_N)=\Z_2$, this is a double cover. In particular, for any loop $\gamma:[0,1]\lto\CC_N$ in the configuration space, there is a corresponding lifted path $\wt{\gamma}:[0,1]\lto\wt{\CC_N}$ in the universal cover whose endpoints are projected to the same point in $\CC_N$. At these points, the wave function should differ only by a sign, leading to constraints:
\begin{align}\label{FR}
    \Psi(\wt{\gamma}(1))=\chi_{\rm FR}(\gamma)\Psi(\wt{\gamma}(0)).
\end{align}
Such constraints \eqref{FR} induced by loops in configuration space are known as Finkelstein--Rubinstein signs \cite{FinkelsteinRubinstein1968}. The sign is determined explicitly by
\begin{align}
    \chi_{\rm FR}(\gamma)=\left\{\begin{array}{cl}
        1 & \gamma\text{ is contractible}, \\
        -1 & \gamma\text{ is not contractible},
    \end{array}\right.
\end{align}
i.e. $\chi_{\rm FR}(\gamma)$ is the representative of $\gamma$ in the homotopy group $\pi_1(\CC_N)$.

As an example, in the rigid body quantization, the relevant loops are those induced by symmetries: every symmetry of a skyrmion gives rise to a constraint of the form
\begin{align}
    \exp\left({\rm i}\alpha \vec{n}\cdot\hat{K}+{\rm i}\beta\vec{N}\cdot\hat{L}\right)\Psi=(-1)^{N_{\alpha,\beta}}\Psi,
\end{align}
where the symmetry is generated by an isorotation/rotation pair $(p(\alpha,\vec{n}),q(\beta,\vec{N}))$ of angles $\alpha$ and $\beta$ around fixed axes $\vec{n}$ and $\vec{N}$ respectively, and here $\hat{K}$ and $\hat{L}$ are the (body-fixed) angular momentum operators in target space and space respective. It is well-known how to compute these types of constraints for certain Skyrme fields, for example those generated from rational maps \cite{krusch2003homotopy}.

When dealing with more complicated configuration spaces, one will invariably encounter loops which do not arise from zero-mode symmetries. For instance, as relevant for the present work, one may construct ADHM data which has internal symmetries of the form \eqref{internal-symmetry} which do not arise simply from the $\SU(2)^2$ action on a single point. In addition, the formulae from \cite{krusch2003homotopy} only apply to rational map skyrmions. So for a complete quantum treatment of instanton-generated-skyrmions, we require a way to determine the Finkelstein--Rubinstein signs for all loops which arise for Skyrme fields generated from instantons; both symmetries of the form \eqref{general-symm} and those internal to the configuration space \eqref{internal-symmetry}.

This problem has recently been resolved \cite{corkharland2024FR}. A loop in the space of ADHM data is a one-parameter family $(L(t),M(t))\in\A_N$, for $t\in[0,1]$ such that
\begin{align}\label{loop-adhm}
    (L(0),M(0))=(L(1)\Omega^{-1},\Omega M(1)\Omega^{-1}),
\end{align}
for some $\Omega\in\O(N)$. This gives rise to a loop $U(t)$ in the space $\CC_N$, and hence a Finkelstein--Rubinstein sign $\chi_{{\rm FR}}\in\{-1,1\}$. For the Skyrme fields generated by a loop satisfying \eqref{loop-adhm}, the sign is
\begin{align}\label{FR-from-ADHM}
    \chi_{\rm FR}=\det\Omega.
\end{align}
In other words, a loop is contractible if and only if the compensating gauge transformation has determinant $1$, i.e. is in $\SO(N)\subset\O(N)$. The proof of this result may be found in \cite{corkharland2024FR}.

The simple formula \eqref{FR-from-ADHM} complements the process outlined in Section \ref{sec:cluster-method} since all symmetries are accounted for along with their compensating gauge transformations. In particular, explicit formulae for the full ADHM data are not needed to determine these signs; the final ADHM data is modeled on diagonal test data \eqref{generic-test-data}, and all generating loops and symmetries (and hence the compensating gauge transformations) are extracted from the symmetries of the test data via \eqref{general-symm} and \eqref{internal-symmetry}.

\section{The 8-skyrmion as two cubic 4-skyrmions}\label{sec:8-sky}
To demonstrate the methods outlined in the previous section, we shall now consider a specific example which showcases all of the different steps required for studying a quantum system in the Skyrme model using instantons.

\subsection{Physical picture and configuration space}

We will consider the nonlinear extension of the lowest frequency vibrational mode of the $N=8$ twisted cube skyrmion. This is also the unstable mode of the untwisted $N=8$ skyrmion. Physically, both 8-skyrmions can be thought of as two cubic 4-skyrmions stacked atop one another. Skyrmions are typically visualised by first plotting an isosurface of constant baryon charge density.
This is coloured to reflect the value of $U(x)$ at that point, based on the Runge colour sphere. The skyrmions are coloured white/black when $U = \pm i \tau_3$ and red,
green and blue when $U = i(\tau_1 \cos(\alpha) + \tau_2 \sin(\alpha))$ and $\alpha$ is $0, 2\pi/3$ and $4\pi/3$ respectively. The interaction of skyrmions can often be intuitively understood using these colours, as explained in \cite{corkhalcrow2022adhm}. For the twisted 8-skyrmion, the cubes are orientated so their touching face has the same colour, and their touching vertices have either the same (twisted) or opposite (untwisted) colour. 

\begin{figure}[h!]
		\includegraphics[width=\textwidth]{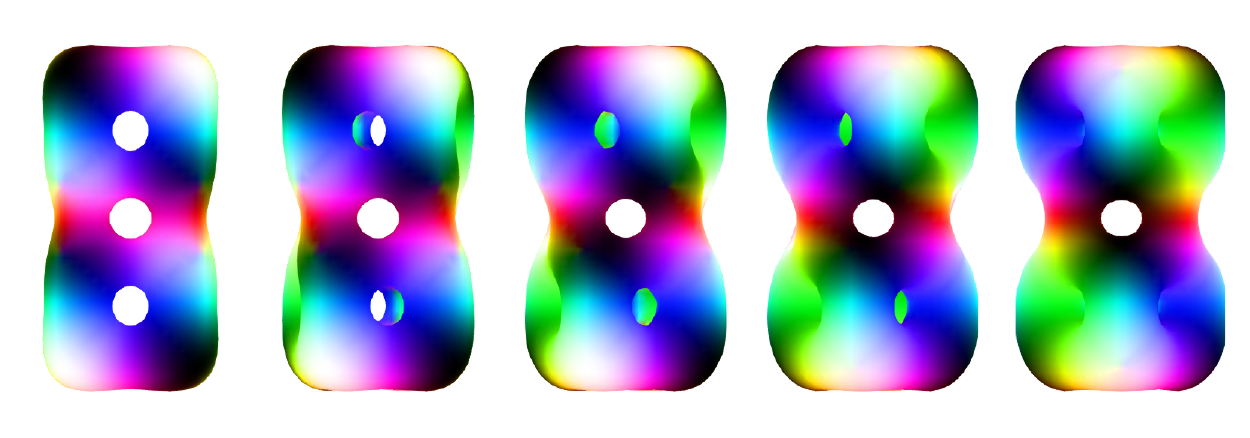}
		\caption{Configurations joining the untwisted (far left) and twisted (far right) $N=8$ skyrmions. The coordinate $\xi$ increases from $0$ to $\pi/4$ from left to right.} \label{fig:config}
\end{figure}

The vibrational mode we consider is the relative rotation of the two cubes around the axis joining them. Let us fix the separation vector along the $z$-axis and parameterise their relative orientation with a coordinate $\xi$. This path is visualised in Figure \ref{fig:config} where we plot several skyrmions for different values of $\xi$ from $0$ to $\tfrac{\pi}{4}$. At $\xi=0$ the cubes have the same orientation, giving the untwisted cubes, while at $\xi=\tfrac{\pi}{4}$ the cubes have different orientations, giving the twisted cubes. 

Due to the symmetry of the cubes, a shift by $\xi\mapsto\xi+\pi$ is a symmetry of the configuration space. In this way $\xi=0$ and $\xi=\pi$ can be identified and so the vibrational manifold has topology of a circle $S^1$. The full configuration space is then the $\SU(2)^2$ bundle over $S^1$ given by acting with rotations and isorotations. $G$-bundles over $S^1$ are classified by connected components of $G$. As $G=\SU(2)^2$ is connected, the configuration space considered here is a trivial (product) bundle $\SU(2)\times\SU(2)\times S^1$.

\subsection{ADHM data for the configuration space}

We now execute the steps described in Section \ref{sec:cluster-method} to generate instantons describing 
the physical picture. 

\textbf{Step 1:} The two clusters used are cubic 4-skyrmions which have zero center-of-mass ADHM data, with scale $\lambda>0$, \cite{LeeseManton1994stable}
\begin{align}\label{data-cube}
\begin{aligned}
L_4&=\lambda\rho\begin{pmatrix}
    \bu&\bi&\bj&\bk
\end{pmatrix},\\
M_4&=\frac{\lambda}{\sqrt{2}}
	\begin{pmatrix}
		0 & -\bj - \bk & -\bi - \bk & -\bi - \bj \\
		-\bj - \bk & 0 & -\bi + \bj & \bi - \bk \\
		-\bi - \bk & -\bi + \bj & 0 & -\bj + \bk \\
		-\bi - \bj & \bi - \bk & -\bj + \bk & 0
	\end{pmatrix}.
\end{aligned}
\end{align}
This data is invariant under the cubic group $\O_h$; more details can be found in \cite{corkhalcrow2022adhm}. The axes of isorotation symmetry depend on the choice of unit quaternion $\rho\in\SU(2)$. We shall fix
\begin{align}
    \rho=\sqrt{\frac{(2-\sqrt{2})(3+\sqrt{3})}{24}}\left(\tfrac{\sqrt{3}-1}{\sqrt{2}}\bu+\bi+(1+\sqrt{2})\bj-\tfrac{1}{2}(\sqrt{3}-1)(\sqrt{2}+2)\bk\right),
\end{align}
which colours each pair of opposite faces of the cube red, green, and blue.

\textbf{Step 2:} To describe the full path, we build diagonal test data from the cubic sub-units as
\begin{align}\label{N=8-test}
    \begin{aligned}
        L_T&=\begin{pmatrix}
            L_4\omega(\xi)^{-1}&L_4\omega(-\xi)^{-1}
        \end{pmatrix}\\
        M_T&=\diag\{\omega(\xi)M_4\omega(\xi)^{-1}+R\id_4\bk,\omega(-\xi)M_4\omega(-\xi)^{-1}-R\id_4\bk\},
    \end{aligned}
\end{align}
where
\begin{align}
    \omega(\xi)=\bu\cos\tfrac{\xi}{2}+\bk\sin\tfrac{\xi}{2}
\end{align}
is the rotation of angle $\xi$ which determines the path of the twist, and $R>0$ is a separation parameter.

\textbf{Step 3:} Every configuration in the configuration space has ${\rm D}_4$ symmetry, generated by
\begin{itemize}
	\item A $\pi$ rotation about $(1,0,0)$.
	\item A $\tfrac{\pi}{2}$ rotation about $(0,0,1)$ and a $\pi$ isorotation about $(1,0,0)$.
\end{itemize}
In the gauge described by \eqref{data-cube}, this symmetry is respected by the test data \eqref{N=8-test}, and understood explicitly by\footnote{Here, and throughout, we employ the notation $p({v},\theta)$ and $q({w},\varphi)$ to denote the unit quaternion representing a rotation by angle $\theta$ or $\varphi$ around the unit axis ${v}$ or ${w}$ respectively. Explicitly
\begin{align*}
    q(w,\varphi)=\cos\tfrac{\varphi}{2}\,\bu+\sin\tfrac{\varphi}{2}\,{w}\cdot(\bi,\bj,\bk)
\end{align*}
and similarly for $p$.}
\begin{align} \label{eq:symms-N=8}
\begin{aligned}
	L_T &= p(e_1,\pi) L_T q(e_3,\tfrac{\pi}{2})^{-1} O_4^{-1}, &M_T&= O_4 q(e_3,\tfrac{\pi}{2}) M_T q(e_3,\tfrac{\pi}{2})^{-1} O_4^{-1}, \\
	L_T &= L_T q(e_1,\pi)^{-1} O_2^{-1},& M_T&= O_2 q(e_1,\pi) M_T q(e_1,\pi)^{-1} O_2^{-1},
\end{aligned}
\end{align}
with compensating gauge transformations
\begin{align}\label{cgts-N=8-symms}
\begin{aligned}
	O_4 &= \begin{pmatrix}  o_4 & 0 \\ 0 & o_4  \end{pmatrix}, &   O_2&= \begin{pmatrix}  0 & o_2 \\ o_2 & 0  \end{pmatrix}, \\
	o_4 &= \begin{pmatrix}0& 0&-1& 0 \\
		-1& 0& 0& 0 \\
		0& 0& 0& 1 \\
		0& -1& 0& 0   \end{pmatrix},&  o_2&= \begin{pmatrix} 0& 1& 0& 0 \\
		-1& 0& 0& 0 \\
		0& 0& 0& -1 \\
		0& 0& 1& 0 \\  \end{pmatrix} .
\end{aligned}
\end{align}
In addition to this ${\rm D}_4$ symmetry, there are internal symmetries induced by moving through the vibrational manifold. Specifically
\begin{itemize}
    \item $\xi\mapsto\xi+\tfrac{\pi}{2}$ and a $\tfrac{\pi}{2}$ rotation about $(0,0,1)$.
    \item $\xi\mapsto-\xi$, a parity inversion, and a $\pi$ isorotation about $(0,0,1)$.
\end{itemize}
The data \eqref{N=8-test} exhibits these symmetries via
\begin{align}\label{ro-vib-symm-B8}
        L_T(\xi)&=L_T(\xi+\tfrac{\pi}{2})q(e_3,\tfrac{\pi}{2})^{-1}O_\sigma^{-1},&M_T(\xi)&=O_\sigma q(e_3,\tfrac{\pi}{2})M_T(\xi+\tfrac{\pi}{2})q(e_3,\tfrac{\pi}{2})^{-1}O_\sigma^{-1}, \nonumber  \\
        L_T(\xi)&=-p(e_3,\pi)L_T(-\xi)O_-^{-1},&M_T(\xi)&=-O_-M_T(-\xi)O_-^{-1},
\end{align}
with compensating gauge transformations
\begin{align}\label{cgt-B8-ro-vib-symm}
\begin{aligned}
    O_\sigma&=\begin{pmatrix}
        o_\sigma&0\\
        0&\id_4
    \end{pmatrix},&O_-&=\begin{pmatrix}
        0&o_-\\
        o_-&0
    \end{pmatrix},\\
    o_\sigma&=\begin{pmatrix}
        0&0&0&1\\
        0&0&-1&0\\
        0&1&0&0\\
        -1&0&0&0
    \end{pmatrix},&o_-&=\frac{\sqrt{3}}{3}\begin{pmatrix}
        0&-1&-1&-1\\
        1&0&-1&1\\
        1&1&0&-1\\
        1&-1&1&0
    \end{pmatrix}.
\end{aligned}
\end{align}

\textbf{Step 4:} The symmetry equations \eqref{eq:symms-N=8} are equivalent to a nullspace problem, which we solve using \texttt{Mathematica}. The most general test data $\mathcal{M}(\boldsymbol{r})$ consistent with the D$_4$ symmetry can be found in Appendix \ref{appendix-D4}.

\textbf{Step 5:}  We now have a set of data $\mathcal{M}(\boldsymbol{r})\in\XX_8$, consistent with D$_4$ symmetry. To become ADHM data these must satisfy the reality condition,
\begin{equation}
    \Im\left(\mathcal{M}(\boldsymbol{r}) \mathcal{M}(\boldsymbol{r})^\dagger \right) = 0 \, ,
\end{equation}
which reduces to nine quadratic equations. The test data \eqref{data-cube} depends on three parameters: $R$, $\lambda$ and $\xi$. The parameters $R$ and $\lambda$ are interpreted as separation and scale, and shall be fixed below; as such the only variable parameter in our configuration space is the coordinate $\xi$. We now want to find the ADHM data closest to the test data by solving \eqref{projection}. We do this using \texttt{Mathematica}'s \texttt{NMinimize} function. We first fix $\xi=\pi/4$ and optimize $\lambda$ and $R$ to minimise the static Skyrme energy. We find the minimiser at $(R,\lambda) = (1.99, 1.142)$; remarkably close to $(R,\lambda) = (2,\sqrt{2})$. Now fixing these two parameters, we find the ADHM data, and hence the skyrmions, for all other $\xi\in [0,\pi/4]$. For each new $\xi$, the initial data is taken as the previous solution. Energy density plots of these configurations are displayed in Figure \ref{fig:config}.

\subsection{Classical results from instanton approximation}

\begin{figure}[h!]
		\includegraphics[width=\textwidth]{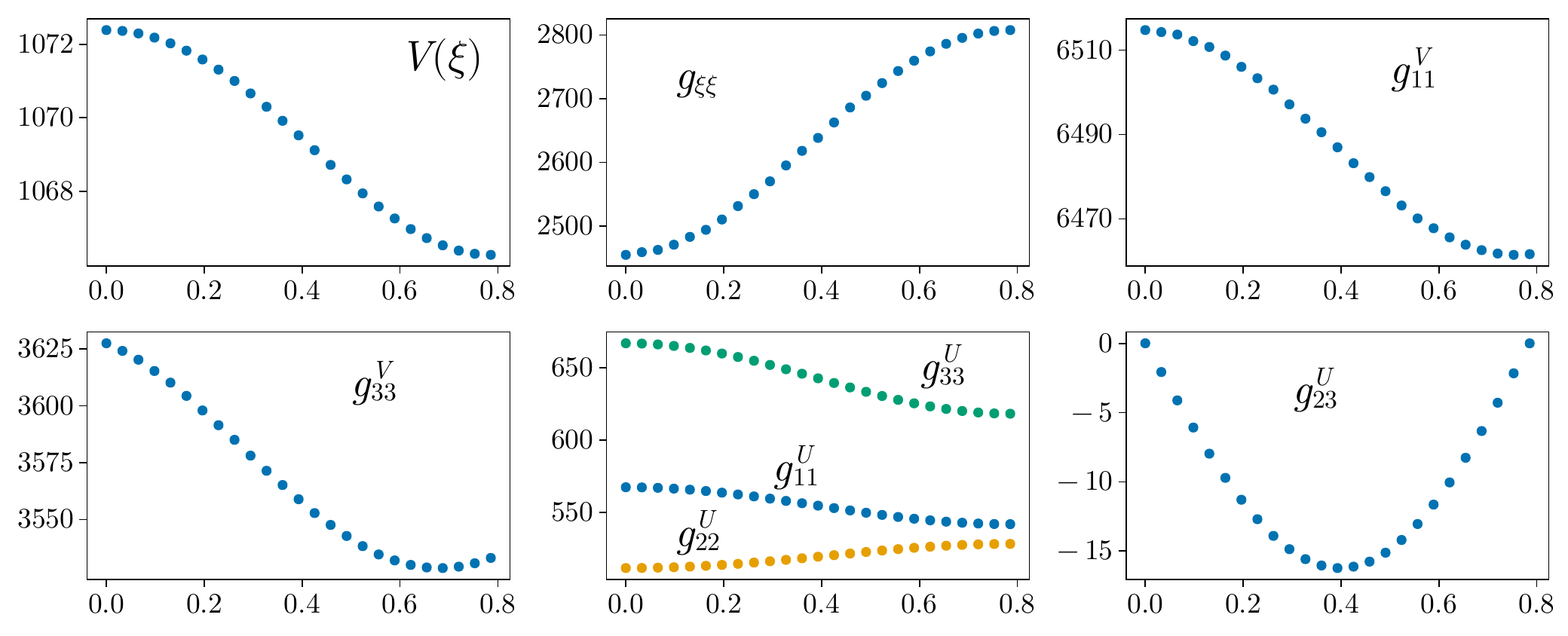}
		\caption{The potential energy and the seven independent metric coefficients, as a function of $\xi$, in Skyrme units. These can be converted to MeV using the factor $f_\pi/4e$. } \label{fig:metric}
\end{figure}

Including rotational and isorotational zero modes, the manifold of configurations is given by $\SU(2)\times\SU(2) \times S^1$. The potential energy and metric on this manifold depend only on the circle, ``vibrational", coordinate $\xi$. We write the metric in terms of moment of inertia coefficients\footnote{Here we use the notation $g^U_{ij},g^V_{ij},g^W_{ij}$ rather than the more standard (e.g. \cite{mankomantonwood2007}) notation $U_{ij},V_{ij},W_{ij}$ so as not to conflate with other notation, and to emphasize that these are metric components.} as
\begin{align}\label{metric-config}
    g=g_{\xi\xi}(\xi)\;\d\xi^2+g_{ij}^U(\xi)\;\alpha_i\alpha_j+g^V_{ij}(\xi)\;\beta_i\beta_j+g^W_{ij}(\xi)\;\alpha_i\beta_j,
\end{align}
where $\alpha_i,\beta_j$ are the one forms dual to the angular velocity vectors \eqref{angular-vel}. We can calculate these coefficients using the formulae in Section \ref{sec:quantizationprocedure}. Note that the cross terms $g_{\xi i}$ vanish because of the parity symmetry \eqref{ro-vib-symm-B8}. Also, because of the D$_4$ symmetry \eqref{eq:symms-N=8}, we find that there are only six independent moment of inertia coefficients. In our orientation, the zero-mode metric tensors take the form
\begin{equation}
	g^U = \begin{pmatrix} g_{11}^U & 0 & 0 \\ 0 & g_{22}^U & g_{23}^U \\ 0 & g_{23}^U & g_{33}^U \end{pmatrix} ,\quad g^V = \begin{pmatrix} g_{11}^V & 0 & 0 \\ 0 & g_{11}^V & 0 \\ 0 & 0 & g_{33}^V \end{pmatrix},  \quad g^W = 0_{3\times 3} \, ,
\end{equation}
where the mixed rotational-vibrational metric terms $g^W$ vanish. We plot the seven independent metric components and the potential energy in Figure \ref{fig:metric}.

\subsection{Quantization}

We'll now use the classical information to construct a model of the Beryllium-8 nucleus. We have constructed the manifold of configurations that have a rotational orientation, isorotational orientation, and a relative twist parameter $\xi$. We'll now implement a canonical quantization on this manifold.

The Schrödinger equation is given by
\begin{equation}
	\left(-\frac{\hbar^2}{2}\Delta_g + V\right)\Psi = E\Psi,
\end{equation}
where $\Delta_g$ is the Laplace--Beltrami operator on $\SU(2)\times\SU(2) \times S^1$ induced by the metric \eqref{metric-config}. Due to rotational-isorotational symmetry, we can write the wavefunction using an expansion in spin states
\begin{equation}
	\Psi = \sum_{L_3 = -J}^J \sum_{K_3 = -I}^I \psi_{L_3,K_3}(\xi) \ket{J, J_3, L_3}\ket{I, I_3, K_3} \, .
\end{equation}
$J \in \mathbb{N}$ is spin, $I \in \mathbb{N}$ is isospin, $J_3/I_3$ are the space-projected spin/isospin and $L_3/K_3$ are the body-projected spin/isospin.  The space-projected spins have no effect on the energy spectrum, so we are free to choose $J_3=I_3=0$ and use the reduced notation
\begin{equation} \label{eq:fullwv}
	\Psi = \sum_{L_3,K_3} \psi_{L_3,K_3}(\xi) \ket{J ,L_3}\ket{I,K_3} \, .
\end{equation}
Only certain wavefunctions are allowed due to the symmetry of the system. All configurations considered are invariant under D$_4$ symmetry, realized as a $\pi$ rotation about $(1,0,0)$ and a  $\tfrac{\pi}{2}$ rotation about $(0,0,1)$ combined with a $\pi$ isorotation about $(1,0,0)$. These give the Finkelstein--Rubinstein constraints
\begin{align}
    \exp\left(i \pi \hat{L}_1\right) \Psi &= (-1)^{N_1} \Psi \label{eq:FR1} \\
    \exp\left(i \tfrac{\pi}{2} \hat{L}_3 + i \pi \hat{K}_1 \right) \Psi &= (-1)^{N_2} \Psi  \label{eq:FR2} \, .
\end{align}
Using the formula \eqref{FR-from-ADHM} from \cite{corkharland2024FR}, we compute that both compensating gauge transformations \eqref{cgts-N=8-symms} have determinant $1$, and so $N_1=N_2=0$. This matches the result expected based on a rigid body analysis \cite{krusch2003homotopy}. The operators that make up these symmetry elements are simple when applied to the spin basis:
\begin{align}
	\exp\left(i \pi \hat{L}_1 \right)\ket{J ,L_3}  \ket{I, K_3} &=  (-1)^{J} \ket{J, -L_3}  \ket{I, K_3} \label{eq:FR1eq}  \\
	\exp\left(i \frac{\pi}{2} \hat{L}_3 +i\pi \hat{K}_1\right)\ket{J, L_3} \ket{I, K_3}  &= i^{L_3 }(-1)^I \ket{J, L_3} \ket{I, -K_3} \label{eq:FR2eq} 
\end{align}
The first FR constraint \eqref{eq:FR1} implies that all rotational wavefunctions must either have even spin and appear as $\ket{J, L_3} + \ket{J, -L_3}$, or odd spin and appear in the combination $\ket{J, L_3} - \ket{J, -L_3}$. The second FR constraint \eqref{eq:FR2} implies that $L_3$ must be even, and that if $I+L_3/2$ is even, the isorotational wavefunctions appear in pairs $\ket{I, K_3} + \ket{I, -K_3}$ while if $I+L_3/2$ is odd the isorotational wavefunctions must appear as $\ket{I, K_3} - \ket{I ,-K_3}$. Overall, the allowed wavefunctions take the form
\begin{equation}
\left( \ket{J, L_3} + (-1)^{J} \ket{J, -L_3} \right)\left(  \ket{I, K_3} + (-1)^{I+L_3/2} \ket{I, -K_3} \right) \, .
\end{equation}
Note that if $J$ is odd there is no allowed state with $L_3=0$ and, similarly, if $I+L_3/2$ is odd there is no allowed state with $K_3=0$. In particular, there are no states with spin-projection $L_3=2$ and isospin $I=0$.

Next, there are combined rotational-vibrational transformations exhibited explicitly via \eqref{ro-vib-symm-B8}. The first is $\xi\mapsto \xi+\pi/2$ combined with a $\pi/2$ rotation about $(0,0,1)$. This transformation provides the third and final FR constraint
\begin{equation}
    \exp\left(i \tfrac{\pi}{2} \hat{L}_3 \right) \Psi(\xi+\tfrac{\pi}{2}) = (-1)^{N_3} \Psi(\xi).\label{eq:FR3}
\end{equation}
This constraint may be determined again using the formula \eqref{FR-from-ADHM} and the determinant of the compensating gauge transformation $O_\sigma$ in \eqref{cgt-B8-ro-vib-symm} corresponding to the symmetry of the test data; this works because the test data is true ADHM data in the limit $R\to \infty$, which can be continuously deformed into the data we use in our work. The determinant is $1$, and as such $N_3=0$. This matches physical intuition: the transformation \eqref{eq:FR3} corresponds to rotating one cube by $\pi$ and leaving the other one untouched. The FR sign of a single cube rotated by $\pi$ is $+1$, matching the result from the new method based on ADHM data.

Finally, there is a parity symmetry. The full parity operator ($U(x) \mapsto U(-x)^{-1}$) can be combined with $\xi \mapsto -\xi$ and a $\pi$ isorotation about $(0,0,1)$. So, whatever our final wavefunction is, we can calculate its parity by the operator
\begin{equation} \label{eq:parityoperator}
    \hat{\mathcal{P}} = \exp\left( i \pi \hat{K}_3 \right) \hat{\mathcal{P}}_\xi \, . 
\end{equation}

It is convenient to now split the symmetries into their (iso)rotational parts and the ``vibrational" parts. The vibrational transformations are just $\xi \mapsto \xi+\pi/2$ and $\xi \mapsto -\xi$. These two operations generate the group C$_{2h}$, which has four irreducible representations: $A_g, B_g, A_u$ and $B_u$. Each vibrational wavefunction, $\psi(\xi)$, transforms as one of these. To satisfy \eqref{eq:FR3} the vibrational wavefunctions which transform trivially under $\xi \mapsto \xi + \pi/2$ (the $A_g$ and $A_u$ wavefunctions) must be combined with the spin states with $L_3 = 0, 4, 8, \ldots$; those which pick up a sign must be combined with the spin states with $L_3 = 2, 6, 10, \ldots$. Finally, the parity of the total wavefunction is given by the combined action \eqref{eq:parityoperator}. Vibrational wavefunctions which transform as $A_g$ or $B_g$ transform trivially under $\xi \mapsto -\xi$, while those which transform as $A_u$ or $B_u$ pick up a sign. Then isospin wavefunctions with even (odd) projection $K_3$ transform trivially (pick up a sign) under the operator $\exp\left( i \pi \hat{K}_3 \right)$. So the combination of these transformations give the overall parity $P$.


The vibrational wavefunctions satisfy Schr\"odinger's equation \eqref{eq:basic-Schro}, which depends on the metric $g$. The results in Figure \ref{fig:metric} show that $g_{23}^U$ is orders of magnitude smaller than the other metric coefficients. Further, $g_{23}^U$ is the only term which mixes the wavefunctions with different $|K_3|$. Hence it is physically reasonable, and theoretically simpler, to set $g_{23}^U=0$ which we do from now on. The D$_4$ symmetry ensures that 
\begin{equation}
	\psi_{L_3,K_3} = \psi_{L_3,-K_3} =\psi_{-L_3,K_3} =\psi_{-L_3,-K_3} \, .
\end{equation}
The symmetry means that each allowed state depends on only one vibrational wavefunction. The Schr\"odinger equation determining the wavefunction is
\begin{multline}
-\frac{\hbar^2}{2\sqrt{g_{\xi\xi}}}\frac{\d}{\d\xi}\left( \frac{1}{\sqrt{g_{\xi\xi}}}\frac{\d \psi(\xi)_{L_3,K_3}}{\d\xi } \right)  + V(\xi)\psi_{L_3,K_3}(\xi) \\ + \frac{\hbar^2}{2} E_{L_3,K_3}(\xi)\psi_{L_3,K_3}(\xi) = E_v \psi_{L_3,K_3}(\xi)
\end{multline}
where
\begin{equation} \label{eq:spinpart}
E_{L_3,K_3}(\xi) =  \bra{I ,K_3}\bra{J, L_3} \left( (\hat{\boldsymbol{L}}, \hat{\boldsymbol{K}})\cdot (g^{R})^{-1}(\xi) \cdot (\hat{\boldsymbol{L}}, \hat{\boldsymbol{K}}) \right) \ket{I, K_3}\ket{J, L_3} \, ,
\end{equation}
and $g^R$ is the (iso)rotational, $6\times 6$ submetric. The metric contains no spin-isospin mixing (i.e. $g^R=\diag\{g^U,g^V\}$), so the spin and isospin parts decouple and can be considered separately. We have that
\begin{equation}
\bra{J, L_3} \hat{\boldsymbol{L}} \cdot (g^V)^{-1} \cdot \hat{\boldsymbol{L}} \ket{J, L_3} = \frac{1}{g_{11}^V}J(J+1) + \left(\frac{1}{g_{33}^V} - \frac{1}{g_{11}^V} \right) L_3^2
\end{equation}
while
\begin{align}
    \bra{1,0} \hat{\boldsymbol{K}}  \cdot (g^U)^{-1} \cdot \hat{\boldsymbol{K}} \ket{1,0} &= \frac{1}{g_{11}^U} + \frac{1}{g_{22}^U} \\
    \left(\bra{1,1} + \bra{1,1} \right) \hat{\boldsymbol{K}}  \cdot (g^U)^{-1} \cdot \hat{\boldsymbol{K}}\left(\ket{1,1} + \ket{1,1} \right) &= \frac{1}{g_{11}^U} + \frac{1}{g_{33}^U} \\
    \left(\bra{1,1} - \bra{1,1} \right) \hat{\boldsymbol{K}}  \cdot (g^U)^{-1} \cdot \hat{\boldsymbol{K}}\left(\ket{1,1} - \ket{1,1} \right) &= \frac{1}{g_{22}^U} + \frac{1}{g_{33}^U} \, .
\end{align}
Finally, we can reinsert physical units and expand the $\xi$ metric term. Overall, the Schr\"odinger equation for each vibrational wavefunction takes the form
\begin{equation} \label{eq:finalSchro}
\frac{f_\pi}{4e}\left(-\frac{2e^4}{g_{\xi\xi}}\left(\frac{\d^2}{\d\xi^2} - \frac{\partial_\xi g_{\xi\xi}}{2 g_{\xi\xi} }  \frac{\d}{\d\xi}\right) +  V(\xi) + 2e^4 E_{L_3, K_3} (\xi)\right)\psi_{L_3,K_3}= E_v \psi_{L_3,K_3} \, .
\end{equation}

\subsection{Results}

\begin{figure}[h!]
\includegraphics[width=\textwidth]{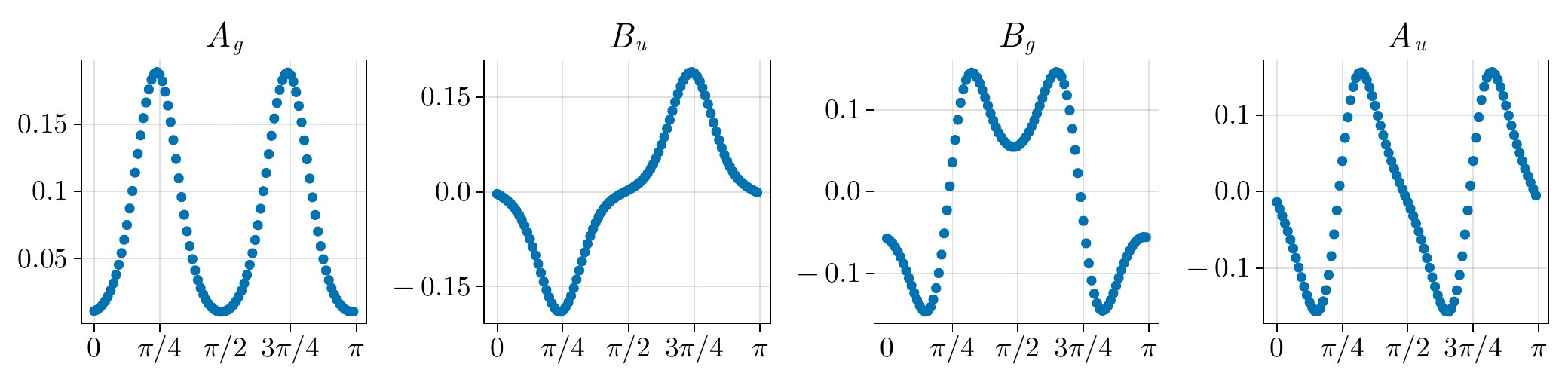}
		\caption{The four lowest energy vibrational wavefunctions, ordered left to right by energy, for $J=I=0$ and $f_\pi = 184$ MeV and $e=3.3$.} \label{fig:wvfns}
\end{figure}

To solve the equation \eqref{eq:finalSchro} numerically, we extend the domain from $[0,\pi/4]$ to $[0,\pi]$. On the extended domain, all the vibrational wavefunctions are periodic and we discretize the Schr\"odinger equation as such, which becomes an eigenvalue problem. The four lowest energy solutions when $J=I=0$ (and hence $E_{L_3, K_3} = 0$) are shown in Figure \ref{fig:wvfns}. When spin or isospin is non-zero, the vibrational wavefunctions receive a small correction. Each vibrational wavefunction is labeled by an irrep, depending on how it transforms under the vibrational group. We remind the reader that wavefunctions which are invariant (pick up a sign) under $\xi \mapsto \xi + \pi/2$ are labeled by $A$ (resp. $B$) and those that are even (odd) under $\xi \mapsto -\xi$ are sub-labeled by $g$ (resp. $u$). Note that the wavefunctions $A_g$ and $B_u$ have similar energies as they are both concentrated at the minimum energy configuration $\xi=\pi/4$. The symmetry of the other wavefunctions forces them to vanish at the minimum, and so they have much higher energy. 

To compare with experimental data, we must fix the physical units of the theory. We choose units to fit the first excited state, with $(I,J) =(0,2)$, to the first experimentally observed state with the same quantum numbers. Hence we take $f_\pi = 184$ MeV and $e=3.3$. The allowed states for each isospin $I=0,1$ and spin $J=0,\dots,4$, and their energies, are displayed in Table \ref{tab:list_of_states}. 
For each spin/isospin structure, there are two states: one with lower energy and one with higher energy. The low energy states arise from the $A_g$ and $B_u$ wavefunctions. These all have rigid-rotor equivalent states in the Skyrme model \cite{battye2009light}. The states with $B_g$ vibrational wavefunction are equivalent to the rigid body states of the untwisted cube. These have a high energy mostly because the untwisted cube has much higher energy than twisted cube. Finally the $A_u$ wavefunction produces states which cannot be described by a rigid rotor, and these have very high energy. 

There are three low-lying states of $^8$Be which form a rotational band and are seen experimentally. Our model also contains many states with very high energy (greater than $25$MeV). As energy increases, so does the number of experimentally observed states. Hence it is impossible to match our states with any observed states. The most important fact is that our model does not introduce new states which are not consistent with experimental data. 

\begin{table}
    \centering
    \begin{tabular}{cc|cccccc}
$I$ & $J$ & Wavefunction & $P$ & $E_v$ & $E_{L_3,K_3}$ & $E$ & \text{Experiment}\\ \hline 
$0$ & $0$ & $\psi_{Ag}\ket{0,0}$ & $+$ & 0.0 & 0.0 & 0.0 & 0.0 \\
 & $0$ & $\psi_{Au}\ket{0,0}$ & $-$ & 36.9 & 0.0 & 36.9 & \\
 & $2$ & $\psi_{Ag}\ket{2,0}$ & $+$ & 0.0 & 3.1 & 3.1 & 3.03 \\
 & $2$ & $\psi_{Au}\ket{2,0}$ & $-$ & 36.9 & 3.1 & 40.0 & \\
 & $4$ & $\psi_{Ag}\ket{4,0}$ & $+$ & 0.0 & 10.2 & 10.2 & 11.35 \\
 & $4$ & $\psi_{Au}\ket{4,0}$ & $-$ & 36.9 & 10.2 & 47.1 & \\
 & $4$ & $\psi_{Ag}(\ket{4,4}+\ket{4,-4})$ & $+$ & 0.0 & 37.4 & 37.4 & \\
 & $4$ & $\psi_{Au}(\ket{4,4}+\ket{4,-4})$ & $-$ & 36.9 & 37.2 & 74.1 & \\\hline
$1$ & $0$ & $\psi_{Au}(\ket{1,1}-\ket{1,-1})\ket{0,0}$ & $+$ & 36.9 & 11.5 & 48.5 & \\
 & $0$ & $\psi_{Ag}(\ket{1,1}-\ket{1,-1})\ket{0,0}$ & $-$ & 0.0 & 11.6 & 11.6 & \\
 & $2$ & $\psi_{Bg}(\ket{1,1}+\ket{1,-1})(\ket{2,2}+\ket{2,-2})$ & $-$ & 35.1 & 17.6 & 52.7 & \\
 & $2$ & $\psi_{Bu}(\ket{1,1}+\ket{1,-1})(\ket{2,2}+\ket{2,-2})$ & $+$ & 0.1 & 17.8 & 17.9 & 16.6 \\
 & $2$ & $\psi_{Bu}\ket{1,0}(\ket{2,2}+\ket{2,-2})$ & $-$ & 0.1 & 18.8 & 18.9 & \\
 & $2$ & $\psi_{Bg}\ket{1,0}(\ket{2,2}+\ket{2,-2})$ & $+$ & 35.1 & 18.8 & 53.9 & \\
 & $2$ & $\psi_{Au}(\ket{1,1}-\ket{1,-1})(\ket{2,0})$ & $+$ & 36.9 & 14.6 & 51.5 & \\
 & $2$ & $\psi_{Ag}(\ket{1,1}-\ket{1,-1})(\ket{2,0})$ & $-$ & 0.0 & 14.7 & 14.7 & \\
 & $3$ & $\psi_{Bg}(\ket{1,1}+\ket{1,-1})(\ket{3,2}-\ket{3,-2})$ & $-$ & 35.1 & 20.7 & 55.8 & \\
 & $3$ & $\psi_{Bu}(\ket{1,1}+\ket{1,-1})(\ket{3,2}-\ket{3,-2})$ & $+$ & 0.1 & 20.9 & 21.0 & 19.1 \\
 & $3$ & $\psi_{Bu}\ket{1,0}(\ket{3,2}-\ket{3,-2})$ & $-$ & 0.1 & 21.9 & 22.0 & \\
 & $3$ & $\psi_{Bg}\ket{1,0}(\ket{3,2}-\ket{3,-2})$ & $+$ & 35.1 & 21.8 & 56.9 & \\
 & $4$ & $\psi_{Au}(\ket{1,1}-\ket{1,-1})\ket{4,0}$ & $+$ & 36.9 & 21.8 & 58.7 & \\
 & $4$ & $\psi_{Ag}(\ket{1,1}-\ket{1,-1})\ket{4,0}$ & $-$ & 0.0 & 21.8 & 21.8 & \\
 & $4$ & $\psi_{Bg}(\ket{1,1}+\ket{1,-1})(\ket{4,2}+\ket{4,-2})$ & $-$ & 35.1 & 24.8 & 59.9 & \\
 & $4$ & $\psi_{Bu}(\ket{1,1}+\ket{1,-1})(\ket{4,2}+\ket{4,-2})$ & $+$ & 0.1 & 25.0 & 25.1 & \\
 & $4$ & $\psi_{Bu}\ket{1,0}(\ket{4,2}+\ket{4,-2})$ & $-$ & 0.1 & 26.0 & 26.1 & \\
 & $4$ & $\psi_{Bg}\ket{1,0}(\ket{4,2}+\ket{4,-2})$ & $+$ & 35.1 & 25.9 & 61.0 & \\
 & $4$ & $\psi_{Au}(\ket{1,1}-\ket{1,-1})(\ket{4,4}+\ket{4,-4})$ & $+$ & 36.9 & 48.7 & 85.7 & \\
 & $4$ & $\psi_{Ag}(\ket{1,1}-\ket{1,-1})(\ket{4,4}+\ket{4,-4})$ & $-$ & 0.0 & 49.0 & 49.0 & 
\end{tabular}
\caption{List of wavefunctions which appear in Figure \ref{fig:EnergySpec}. We list their isospin $I$, spin $J$, spin structure, parity $P$, vibrational energy $E_v$, spin energy $E_{I,J}$, total energy $E$ and (if applicable) the energy of an experimentally observed state which they model.}
\label{tab:list_of_states}
\end{table}

\begin{figure}[h!]
\includegraphics[width=\textwidth]{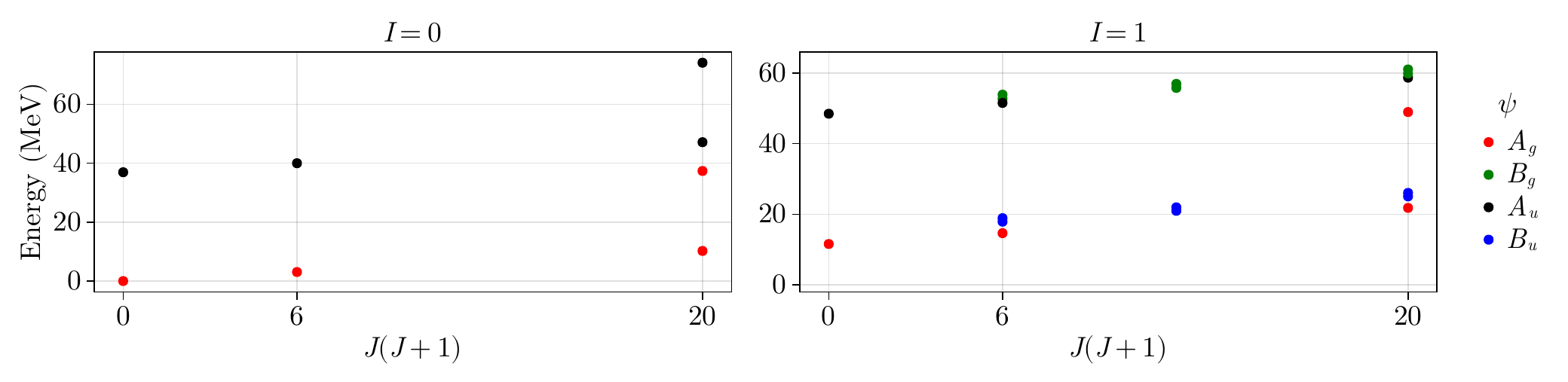}
		\caption{The energy spectrum for our model of Beryllium-8.} \label{fig:EnergySpec}
\end{figure}

The energy spectrum is plotted in Figure \ref{fig:EnergySpec}. In a harmonic, rigid rotor approximation each classical solution gives rise to a set of quantum states called bands, whose energy increases with $J(J+1)$. Figure \ref{fig:EnergySpec} clearly reveals that a similar result occurs in our model but with the bands built upon vibrational wavefunctions. In the $I=0$ plot, we see a band in red, built upon the $A_g$ wavefunction, and a higher energy band in green built upon the $B_g$ wavefunction. In the $I=1$ plot there are three bands, built upon the lowest energy $B_u$, $A_u$ and $B_g$ wavefunctions.

Beryllium-8 is a difficult nucleus to study both theoretically and experimentally, as it is unstable to decay into two $\alpha$-particles. This fact alone motivates models which describe the nucleus as two $\alpha$-particles. The experimental spectrum includes spin $0^+$, $2^+$, and $4^+$ states whose energies are consistent with a quantized rigid rotor. This provides further evidence of the basic model: two $\alpha$-particles in a central potential. This was proposed and studied in the original $\alpha$-particle paper from 1937 \cite{wheeler1937molecular} and the basic picture still holds in modern, more complicated, frameworks such as the algebraic cluster model \cite{bijker2020cluster}.

Our results at isospin 1 are more novel. Boron-8, Beryllium-8 and Lithium-8 form an isospin 1 isotriplet. At the level of energy spectra, the isospin symmetry implies that our isospin $1$ states should appear in all three nuclei spectra. In particular, Boron-8 and Lithium-8 should have approximately equal spectra. However, there are some contradictions in experiments, where many more states are seen for Boron-8 than Lithium-8 \cite{yamaguchi2009low}. Spin $2^+, 1^+$ and $3^+$ states are seen in both, and low energy states with these spins are predicted by the Skyrme model. However, we also predict a low-lying spin $0^-$ state which is not seen for either nucleus. The shell model also predicts a state with spin-parity $0^-$ at around $5$ MeV (fig 2 of  \cite{knox1987reactions}; ``non normal parity" refers to negative parity in this paper). Both our model and the shell model in \cite{knox1987reactions} predict a low energy spin $4^-$ state (our state with energy $21.8$ MeV, $10.2$ MeV above the lowest energy isospin-1 state) which has not been seen for either nucleus. The paper \cite{yamaguchi2009low} reports the existence of a spin $1^-$ state in Boron-8. There are also $1^{\pm}$ states in Beryllium-8. However, our model does not contain any states with this spin. To construct these we would need to include a different vibration. A good candidate is the low-frequency $^2E_u$ vibration described in \cite{GudnasonHalcrow2018vibrational}; a similar state leads to spin $1$ states in a model of Carbon-12 \cite{Rawlinson:2017rcq}.

\subsection{Improving the instanton approximation}

The instanton approximation provides the correct global picture. That is: it produces a manifold of skyrmions with the correct symmetries and topology. However, the energies and metric components are not accurate when compared with true solutions of the Skyrme model. These are known for the twisted and untwisted skyrmions. For zero pion mass $m=0$, the twisted solution the true energy is $E^t = 1037$ while the instanton energy is $E^i=1066$. The metric terms are even worse. For example, the true value of the rotational moment of inertia is $(g_{11}^V)^t = 4458.7$ while the instanton approximation gives $(g_{11}^V)^i = 1.45, (g_{11}^V)^t = 6461.6$. Further, it is believed that a non-zero pion mass is needed to accurately describe real physics \cite{mankomantonwood2007}. The pion mass is modeled by including the term
\begin{equation}
    -\frac{m_\pi^2f_\pi^2}{8\hbar^3}\, \text{tr}( \text{Id}-U)
\end{equation}
in the lagrangian \eqref{eq:SkyrmeLagrangian}. The dimensionless Skyrme energy \eqref{Skyrme-energy} receives the contribution
\begin{equation}
   m^2\,\text{tr}(\text{Id}-U), \quad \text{where} \quad m=\frac{2m_\pi}{f_\pi e} \, .
\end{equation}
We include these terms and compute the twisted and untwisted skyrmions for pion mass $m=1$. The energy and independent moment of inertia terms are listed for the instanton approximation, $m=0$ skyrmions and $m=1$ skyrmions in Table \ref{tab:instanton_vs_nums}.

\begin{table}[h!]
    \centering
    \begin{tabular}{l|cccccc}
Untwisted ($\xi=0$)& $E$ & $ g_{11}^U $ & $g_{22}^U$ & $ g_{33}^U $ & $ g_{11}^V $ & $ g_{33}^V $ \\ \hline
Instanton & 1072.4 & 567.3 & 511.4 & 666.9 & 6514.8 & 3627.5 \\
$m = 0 $ & 1042.2 & 560.1 & 515.7 & 689.0 & 6921.6 & 3288.6 \\
$m = 1$ & 1210.0 & 289.9 & 282.6 & 346.4 & 4384.9 & 1677.2 \\
Twisted ($\xi=\tfrac{\pi}{4}$)&  &  &  & &  &\\ \hline
Instanton & 1066.3 & 541.8 & 528.1 & 618.2 & 6461.6 & 3533.1 \\
$m = 0 \,\, ({\rm D}_{6d})$ & 1037.2 & 589.4 & 510.8 & 578.0 & 4458.7 & 4625.0 \\
$m = 1$ & 1206.7 & 295.3 & 287.0 & 321.6 & 3825.2 & 1684.0 \\
    \end{tabular}
    \caption{Energies and metric coefficients for the twisted and untwisted states given by the instanton approximation (at pion mass $m=0$), and the numerical solutions with pion mass $m=0$ and $m=1$.}
    \label{tab:instanton_vs_nums}
\end{table}

However since the Skyrme properties are understood at special points, we can use these to improve the approximation. One way to do this is to assume that the shape of the potential and metric functions from the instanton approximation is correct, but that the overall zero-point energy and scale are wrong. So we can calibrate around the true values and define a new shifted and scaled potential energy
\begin{equation}
	V^t(\xi) =  \frac{1}{V^i(0) - V^i(\tfrac{\pi}{4})}\left( (V^t(0) - V^t(\tfrac{\pi}{4})) V^i(\xi) + V^t(\tfrac{\pi}{4})V^i(0) - V^t(0)V^i(\tfrac{\pi}{4}) \right),
\end{equation}
with analogous expressions for the metric terms; here superscript $t$ refers to the true values dictated by numerical solutions, and superscript $i$ are the instanton-generated values.

\begin{figure}[h!]
\includegraphics[width=\textwidth]{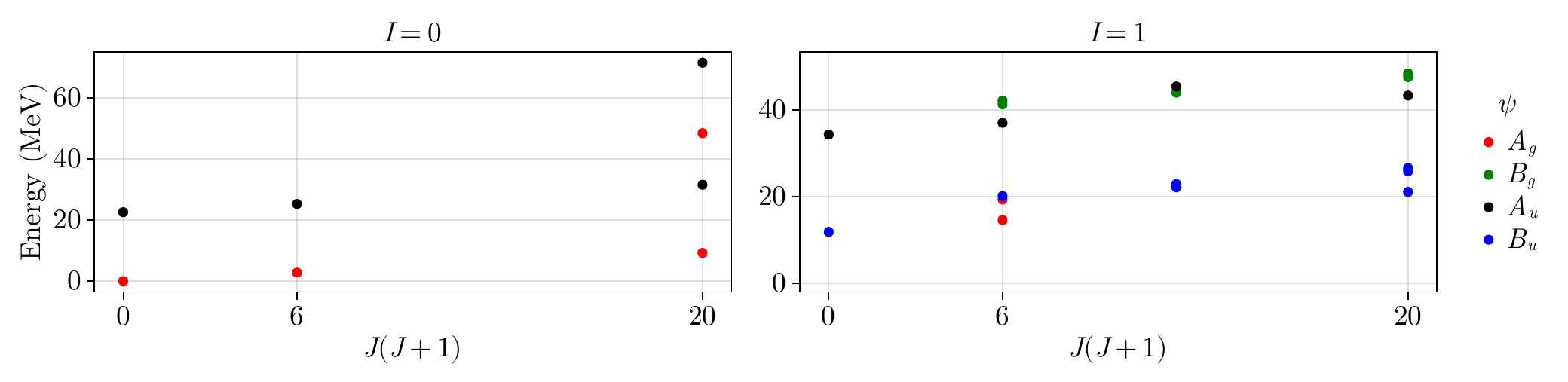}
		\caption{The energy spectrum for the rescaled $m=1$ input.} \label{fig:EnergySpecm1}
\end{figure}

We find the new potential and metric functions and solve the Schr\"odinger equation again using them. The updated spectrum is displayed in Figure \ref{fig:EnergySpecm1}. We adjust the parameters to, once again, fit our first excited state with the first experimental state. Here, $f_\pi = 184$ MeV and $e=2.7$. The final spectrum is very similar to the previous one. Hence the overall effect of rescaling was undone by choosing new physical parameters. So here, the poor match between numerically generated skyrmions by instanton-generated skyrmions is entirely fixed by rescaling physical parameters. This is reassuring for those who want to use instantons to model skyrmion configuration spaces.

Another way to improve on the instanton approximation is to use a string, or elastic band, method. Here, the instanton-generated skyrmions are used as initial data which parameterise an ``elastic band" of configurations. The entire band is then flowed to minimize its total energy while keeping the configurations separated. This was done recently for point-like skyrmions \cite{SpeightWinyard2023nudged}, and the instanton method provides a way to generalise this work to the full Skyrme model.

\section{Paths with no symmetry}\label{sec:paths-no-symm}

In the previous sections, we have been very careful to consider the symmetries and use all available information to construct our ADHM data. In this section, we will be significantly less careful with symmetries but show that the same methods as before can be used to generate interesting paths in the Skyrme configuration space. These paths could be used as initial data in an elastic band calculation \cite{SpeightWinyard2023nudged}. We will follow the steps of Section \ref{sec:cluster-method}, but ignore considerations of gauge fixing and symmetries. That is, we will ignore steps 3 and 4.

\begin{figure}[ht!]
		\includegraphics[width=\textwidth]{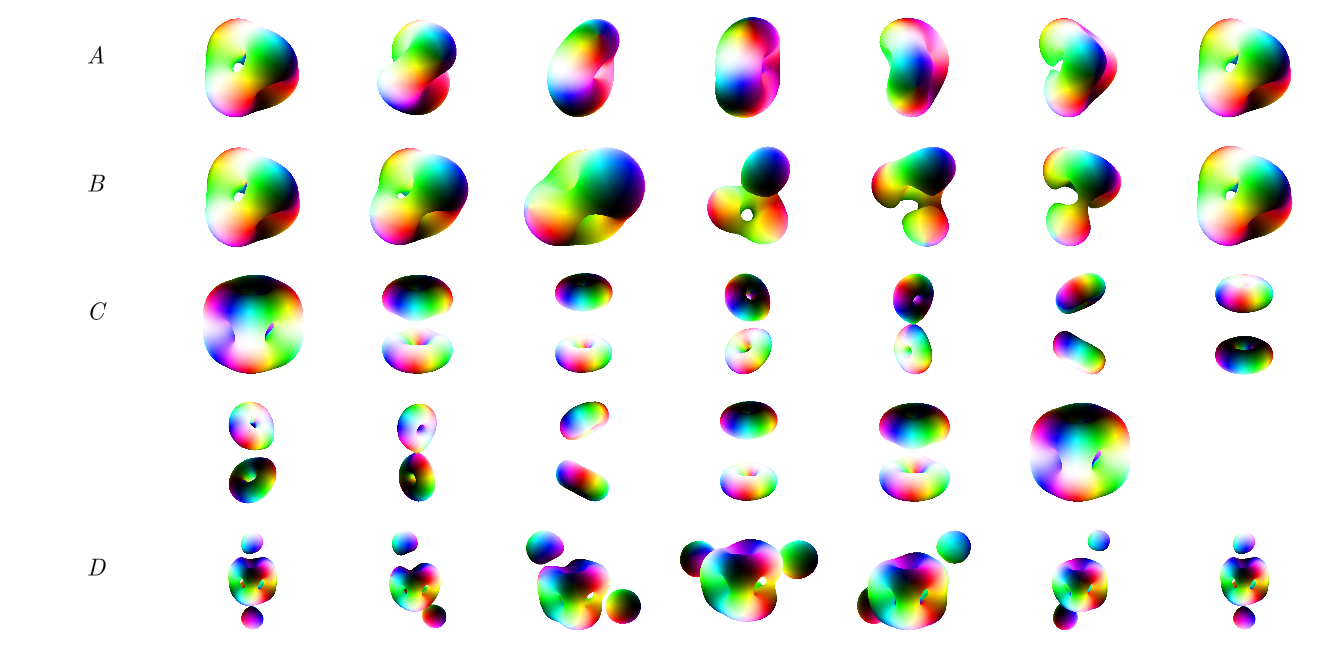}
		\caption{Families of instanton-generated skyrmions, made from the ADHM data closest to the test data defined in this section. All test data are defined with respect to the same parameter $t$, which varies from $0$ to $1$ as the figure goes from left to right. A: test data defined in \eqref{eq:testdataA}. B: test data defined in \eqref{eq:testdataB}. C: test data defined in \eqref{eq:N4split}. D: test data defined in \eqref{eq:N6test}.} \label{fig:lotsa_configs}
\end{figure}

First, consider two paths in the charge 3 sector. We can write the tetrahedral skyrmion using the ADHM data
\begin{equation}
    L = \lambda \begin{pmatrix} 
     \bu &  \bi &  \bk \end{pmatrix} \, \quad
M = \begin{pmatrix}    R \bk & c \bi + c \bj & 0 \\
    c \bi + c\bj & 0 & c \bj - c \bi \\
    0 & c \bj - c \bi & -R \bk
    \end{pmatrix} \, ,
\end{equation}
with $R = \lambda$ and $c = \lambda/\sqrt{2}$. In this gauge the data can be interpreted as three 1-skyrmions with positions $z=R, 0, -R$. We will construct two paths which permute the three skyrmions, starting and ending at the tetrahedron. In the first, we permute the first and third skyrmions, leaving the central one stationary. One such possible test data set is given by
\begin{align} \begin{aligned}\label{eq:testdataA}
L_A(t) &= \lambda \left( \cos(\tfrac{\pi}{2}t)\bu + \sin(\tfrac{\pi}{2}t) \bk \quad \bi \quad  \cos(\tfrac{\pi}{2}t) \bk + \sin(\tfrac{\pi}{2}t) \bu \right), \\ 
M_A(t) &= R\,\text{diag} \left\{  \cos(\pi t) \bk +  \sin(\pi t) \bi, 0 , -\cos(\pi t) \bk -  \sin(\pi t)\bi  \right\}.\end{aligned}
\end{align}
We find ADHM data from this diagonal data by finding the closest data as defined in \eqref{projection}. The skyrmions generated from this data, with $R=\lambda=1$, are shown in Figure \ref{fig:lotsa_configs}A. The initial and final states of the path are related by
\begin{equation}
    L_A(1) = L_A(0)\Omega^{-1},\quad M_A(1) = \Omega M_A(0) \Omega^{-1},\quad \Omega=\begin{pmatrix} 0 & 0 & 1 \\ 0&  1&  0 \\ 1 & 0 & 0 \end{pmatrix}.
\end{equation}
The compensating gauge transform $\Omega$ has determinant $-1$ and hence this path is non-contractible. If this path were included in a quantization, the wavefunction would pick up a sign under this transformation. Further, Morse theory guarantees that a saddle point exists on the energy-minimised path. The true saddle configuration can be found using elastic band methods, as was done recently in a point-particle Skyrme model \cite{SpeightWinyard2023nudged}. The plots of the energy density suggest that the saddle point will be the 3-torus.

A set of test data permuting all three skyrmions is given by
\begin{align}\label{eq:testdataB}\begin{aligned}
L_B(t) &=  \lambda \left(\cos(\tfrac{\pi}{2}t) \bu + \sin(\tfrac{\pi}{2}t) \bk \quad  \cos(\tfrac{\pi}{2}t)\bi + \sin(\tfrac{\pi}{2}t) \bu \quad \cos(\tfrac{\pi}{2}t) \bk + \sin(\tfrac{\pi}{2}t) \bi \right), \\ 
M_B(t) &= R\,\, \text{diag} \left\{ \cos(\pi t) \bk +  \sin(\pi t) \bi,  \sin^2(\tfrac{\pi}{2}t)\bk - \sin(2\pi t)\bi/2  , -\cos(\tfrac{\pi}{2}t)\bk  \right\}.\end{aligned}
\end{align}
The skyrmions generated from the closest data to the diagonal data are plotted in Figure \ref{fig:lotsa_configs}B. The compensating gauge transformation relating the end-points is $\begin{pmatrix}
    0&1&0\\
    0&0&1\\
    1&0&0
\end{pmatrix}$, which has determinant $1$, showing that this loop is contractible.

Now consider a path which splits the cubic 4-skyrmion into two 2-tori, rotates them both and then recombines them into the cube. First consider the ADHM data with cubic symmetry, in a gauge which emphasises the 2-tori structure \cite{corkhalcrow2022adhm},
\begin{align}
    \begin{aligned}
    L_4&=\lambda \begin{pmatrix}
    \bi L_2 & L_2
    \end{pmatrix}, \quad 
    M_4=\begin{pmatrix}
    \mu_1 m_{T^2} + R \bk & \mu_2 m_{\perp} \\
     \mu_2 m_{\perp}^T & \mu_1 m_{T^2} - R\bk 
    \end{pmatrix} \\
    L_2 &= \begin{pmatrix} \bu & \bk \end{pmatrix}, \quad m_{T^2} = \begin{pmatrix} \bi & \bj \\ \bj & -\bi \end{pmatrix}, \quad m_{\perp} = \begin{pmatrix} \bi & -\bj \\ -\bj & -\bi \end{pmatrix}
    \end{aligned}\label{two-tori}
\end{align}
with $\mu_1 = \lambda\sqrt{3}/2$, $\mu_2 = \lambda/2$ and $R=\lambda$. We will take $\lambda=1$ for our simulations. Test data which looks like two clusters being split, rotated, and recombined is given by
\begin{align} \label{eq:N4split}
\begin{aligned}
    L_C&=\begin{pmatrix}
    \bi L_2 q_1(t)^{-1}& L_2 q_2(t)^{-1}
    \end{pmatrix}\\
    M_C&=\diag \left\{
    \mu_1 q_1(t) m_{T^2}q_1(t)^{-1} + R(t)\id_2 \bk,\mu_1 q_2(t) m_{T^2}q_2(t)^{-1} - R(t)\id_2\bk 
    \right\}
\end{aligned}
\end{align}
where $R(t)$ grows linearly from 1 to 2 as $t$ grows from $0$ to $1/3$, remains constant for $t\in[1/3,2/3]$ and decreases from 2 to 1 as $t$ increases from $2/3$ to $1$. The rotation functions $q_1$ and $q_2$ apply rotations around the $x$- and $y$-axes respectively from 0 to $2\pi$ as $t$ grows from $1/3$ to $2/3$ and are unity otherwise. The skyrmions generated from the ADHM data closest to \eqref{eq:N4split} are shown in Figure \ref{fig:lotsa_configs}C.

Finally, consider two 1-skyrmions orbiting a 4-skyrmion. The test data is given by
\begin{align}
\begin{aligned}\label{eq:N6test}
L_D &= \begin{pmatrix}
    L_4& (\bu + \bk)q(\boldsymbol{e}_2, \pi t)^{-1} &  q(\boldsymbol{e}_3, \pi t)(\bi - \bj) q(\boldsymbol{e}_1, \pi t)^{-1} 
\end{pmatrix} \\
M_D &= \diag\left\{
    M_4,  q({e}_2, \pi t) (R\bk) q({e}_2, \pi t)^{-1},  q({e}_1, \pi t)  (-R\bk) q({e}_1, \pi t)^{-1} 
\right\}.
\end{aligned}
\end{align}
As $t$ varies from $0$ to $1$, the two 1-skyrmions will change places, and swap their orientations. Hence this is a loop, and since the compensating gauge transformation has determinant $-1$, we find that it is non-contractible. The skyrmions generated from the ADHM data closest to \eqref{eq:N6test} are shown in Figure \ref{fig:lotsa_configs}D. In this case, the results were better if we only included a subset of elements in the algorithm to calculate the closest data. We only included the $4\times4$ block and the $2\times 2$ block containing the two 1-skyrmions when calculating the distance between the test data and the ADHM data.

In this Section we have seen four examples of creating paths in the Skyrme configuration space from simple test data. We have shown that without much analytical effort, one can create many interesting paths. The paths created are not necessarily usable themselves; the configurations likely have very high energy. Instead, they will help construct initial paths for numerical schemes such as those used in \cite{SpeightWinyard2023nudged}.

\section{Conclusion}

We have described a step-by-step procedure for how to construct manifolds of instanton-generated skyrmions using ADHM data. These manifolds may be used to study the space of all Skyrme fields; most importantly, to semiclassically quantize the model and study nuclei. Following our own procedure, we generated a particular space which described the nonlinear extension of the lowest vibrational mode of the untwisted 8-skyrmion. Quantization on this space produces a model of Beryllium-8. We applied a new method to calculate the Finkelstein--Rubinstein constraints on the space, and compared the results of our model to data. 

It has been long-known that instantons provide an excellent approximation of skyrmions, being the only approximation that can describe skyrmions when they are well separated and when they are coalesced into a high-symmetry object. However, there have been few applications in practice \cite{Leese:1994hb}. We hope this paper will serve as motivation and a template for further work. 

Each $N$-skyrmion has approximately $7N$ normal modes \cite{GudnasonHalcrow2018vibrational}. Each of these has a nonlinear extension and should be treatable in a similar way to the mode discussion in Section \ref{sec:8-sky}. Hence there is a deep well of problems for future work. It would be interesting to develop general methods to describe normal modes in the instanton approximation building on past work for $N=3$ \cite{houghton1999-3skyrme}, although there are subtleties relating to the instanton position in the fourth, ``holographic" direction \cite{Halcrow:2021wwc}. But the more difficult problem is to realise which modes should be treated carefully and which are well-described harmonically. In this paper, we studied a mode linking two low-energy solutions and whose nonlinear extension is compact. Neither of these facts are included in a harmonic approximation.

As well as our step-by-step method, we considered a more careless ad-hoc approach for generating skyrmions from instantons. Here, we used a numerical method to find the ADHM data closest to some test data, without fixing symmetries. This allowed us to quickly generate paths of skyrmions. These could be used as initial data for elastic band methods \cite{SpeightWinyard2023nudged} or simply as a tool to get qualitative insights about the skyrmion configuration space.

\section*{Acknowledgments}
We thank Nick Manton for clarifying discussions about parity and comparisons with other nuclear models. CH was supported by the Carl Trygger Foundation through the grant CTS 20:25.

\appendix

\section{General \texorpdfstring{D$_4$}{D4} symmetric data} \label{appendix-D4}

The most general $(L, M)$ that is consistent with the symmetries \eqref{eq:symms-N=8} takes the form
\begin{align*}
L =&\big( \bu s+\bi t+\bk (-s-\sqrt{2} u)+\bj (t+\sqrt{2} v) ,
\bj (-\sqrt{2} s-u)+\bi u+\bk (-\sqrt{2} t-v)-\bu v,\\
&\bj u+\bi (\sqrt{2} s+u)+\bu (-\sqrt{2} t-v)+\bk v,
\bk s-\bj t+\bu (s+\sqrt{2} u)+\bi (t+\sqrt{2} v) ,\\
&\bk (-\sqrt{2} s-u)+\bu u+\bi v+\bj (\sqrt{2} t+v),
\bi s-\bu t+\bj (-s-\sqrt{2} u)+\bk (-t-\sqrt{2} v),\\
&\bj s+\bk t+\bi (s+\sqrt{2} u)+\bu (-t-\sqrt{2} v),
\bk u+\bu (\sqrt{2} s+u)-\bj v+\bi (\sqrt{2} t+v) \big) \, 
\end{align*}
and
\begin{equation}
    M = \begin{pmatrix} M_{11} & M_{12} \\ M_{12}^t & M_{22} \end{pmatrix} \, ,
\end{equation}
with
\begin{align*}
&M_{11} = \\
&\begin{pmatrix}
\bu a+\bi b+\bj c+\bk d & 
\bu e+\bi f+\bj g+\bk h & 
\bu e-\bj f+\bi g+\bk h & 
\bi i+\bj j
\\
\bu e+\bi f+\bj g+\bk h & 
\bu a+\bj b-\bi c+\bk d & 
-\bj i+\bi j & 
\bu e+\bj f-\bi g+\bk h
\\
\bu e-\bj f+\bi g+\bk h & 
-\bj i+\bi j & 
\bu a-\bj b+\bi c+\bk d & 
-\bu e+\bi f+\bj g-\bk h
\\
\bi i+\bj j & 
\bu e+\bj f-\bi g+\bk h & 
-\bu e+\bi f+\bj g-\bk h & 
\bu a-\bi b-\bj c+\bk d
    \end{pmatrix} \\
&M_{12} = 
    \begin{pmatrix}
\bu k+\bi l-\bj l & 
\bj m+\bk n & 
\bi o+\bk p & 
\bu q+\bi r+\bj r
\\
\bj o+\bk p & 
\bu k+\bi l+\bj l & 
-\bu q+\bi r-\bj r & 
-\bi m+\bk n
\\
\bi m+\bk n & 
\bu q+\bi r-\bj r & 
\bu k-\bi l-\bj l & 
\bj o-\bk p
\\
-\bu q+\bi r+\bj r & 
-\bi o+\bk p & 
\bj m-\bk n & 
\bu k-\bi l+\bj l
    \end{pmatrix}\\
&M_{22} = \\
&\begin{pmatrix}
 \bu a-\bj b-\bi c-\bk d & 
-\bu e-\bi f+\bj g+\bk h & 
-\bu e+\bj f+\bi g+\bk h & 
\bj i+\bi j
\\
-\bu e-\bi f+\bj g+\bk h & 
\bu a+\bi b-\bj c-\bk d & 
\bi i-\bj j & 
-\bu e-\bj f-\bi g+\bk h
\\
-\bu e+\bj f+\bi g+\bk h & 
\bi i-\bj j & 
\bu a-\bi b+\bj c-\bk d & 
\bu e-\bi f+\bj g-\bk h
\\
\bj i+\bi j & 
-\bu e-\bj f-\bi g+\bk h & 
\bu e-\bi f+\bj g-\bk h & 
\bu a+\bj b+\bi c-\bk d
    \end{pmatrix} \, .
\end{align*}
\newpage
\bibliographystyle{unsrt}
\bibliography{refs}
\end{document}